\documentclass[letterpaper, 10 pt, conference]{ieeeconf}  

\IEEEoverridecommandlockouts                              

\usepackage{cite}
\usepackage{hyperref}
\let\labelindent\relax
\usepackage{enumitem}
\usepackage{amsmath,amssymb,amsfonts}
\usepackage{subcaption}
\usepackage{mathtools}
\usepackage{algorithm}
\usepackage{algpseudocode}
\usepackage{stfloats} 
\usepackage{graphicx}
\usepackage{textcomp}
\usepackage{xcolor}

\newcommand{\La}{\mathcal{L}}
\graphicspath{{figures/}} 
\def\BibTeX{{\rm B\kern-.05em{\sc i\kern-.025em b}\kern-.08em
    T\kern-.1667em\lower.7ex\hbox{E}\kern-.125emX}}

\newcommand{\Mplus}{\mathbb{M}^{n}_{+}}      
\newcommand{\E}{\mathbb{E}}
\newcommand{\ip}[2]{\langle #1,\,#2\rangle}

\newcommand{\Proj}{\Pi}
\newcommand{\RR}{\mathbb{R}}
\newcommand{\qedhere}{\hfill$\square$}

\newtheorem{assumption}{Assumption}
\newtheorem{lemma}{Lemma}
\newtheorem{proposition}{Proposition}
\newtheorem{theorem}{Theorem}
\newtheorem{remark}{Remark}

\begin{document}

\title{Asymmetric regularization mechanism for GAN training with Variational Inequalities}


\author{Spyridon C.~Giagtzoglou 
\and
Mark H.M. Winands
\and
Barbara Franci
\thanks{
S. C. Giagtzoglou was supported by the Dutch Research Council
(NWO) M 1 grant, project number OCENW.M.22.433. S. C. Giagtzoglou and Mark H.M. Winands are with the Department of Advanced Computing Sciences, Maastricht University, Paul-Henri Spaaklaan 1, 6229 GS Maastricht, The Netherlands (spyridon.giagtzoglou@maastrichtuniversity.nl; m.winands@maastrichtuniversity.nl).
B. Franci is with the Department of Mathematical Sciences, Politecnico di Torino, Corso Duca degli Abruzzi 24, 10129 Torino, Italy (barbara.franci@polito.it).}}

\maketitle

\begin{abstract}
We formulate the training of generative adversarial networks (GANs) as a Nash equilibrium seeking problem. To stabilize the training process and find a Nash equilibrium, we propose an asymmetric regularization mechanism based on the classic Tikhonov step and on a novel zero-centered gradient penalty. Under  smoothness and a local identifiability condition induced by a Gauss--Newton Gramian, we obtain explicit Lipschitz and (strong)-monotonicity constants for the regularized operator. These constants ensure last-iterate linear convergence of a single-call Extrapolation-from-the-Past (EFTP) method. Empirical simulations on an academic example show that, even when strong monotonicity cannot be achieved, the asymmetric regularization is enough to converge to an equilibrium and stabilize the trajectory.
\end{abstract}

\section{Introduction}
\label{sec:intro}

Generative Adversarial Networks (GANs) are a machine-learning model composed of two neural networks: the discriminator and the generator\cite{goodfellow2014generativeadversarialnetworks}. The purpose of the model is to train the generator to create realistic data (images, audio) by playing a ``game" against the discriminator. In particular, the training of the two neural networks is a two-player zero-sum game that can inherently be modelled as a saddle-point problem.
Saddle point problems are well studied in the literature \cite{li2025primal,facchinei2003finite} and find applications in consensus problems, resource allocation or simply primal-dual problems \cite{goldsztajn2020proximal, HuangTAC2024}. However, GANs highlighted some instabilities, like mode collapse, that can be seen as “best-response” pathologies. Moreover, the training dynamics are often rotational and poorly conditioned and the only available solutions 
are basically ways to change the game’s payoff to make the learning process smoother \cite{gulrajani2017improvedtrainingwassersteingans, jolicoeurmartineau2018relativisticdiscriminatorkeyelement}.

We rely instead on variational inequalities (VIs) and monotone operator theory to establish the properties of the operators involved and derive projected algorithms with convergence guarantees.
We adopt a VI viewpoint because it exposes the saddle operator connected to the GAN game and its geometry, it allows to handle constraints via projections in a principled way, and directly connects to convergence guarantees for a variety of methods. 
Since the VI formulation highlights the rotational dynamics of adversarial training, stabilization strategies are needed to make the training reliable.
For this reason, we introduce a regularization mechanism applied \emph{only on the discriminator side}, that is, asymmetrically across the saddle blocks. This discriminator-only block-asymmetric regularization adds curvature where it is most needed while preserving the target equilibrium point. It also provides explicit analytic constants, specifically Lipschitz continuity and strong-monotonicity constants for the regularized operator, thereby enabling convergence guarantees for the proposed iterative method. 

Among the available algorithms in the literature, we propose Extrapolation-from-the-Past (EFTP) \cite{Popov1980AMO} because, although made of two projection steps, it requires only one gradient computation, hence providing a computationally convenient alternative to the more common Forward-Backward (FB) \cite{FranciGrammatico2022} and ExtraGradient (EG)\cite{Korpelevich1976} schemes.
In particular, our contributions can be summarized as follows.
\begin{itemize}
    \item We introduce and analyze two explicitly \emph{asymmetric} regularizations based on Tikhonov \cite{Tikhonov1995} and a zero-centered input-gradient penalty on the discriminator side. The penalty vanishes at equilibrium, preserving the original saddle point structure while improving conditioning.
    \item Under smoothness and a local Gauss--Newton identifiability condition, we derive explicit Lipschitz and strong-monotonicity constants for the regularized operator, showing that our regularization can infer stronger monotonicity properties on the original saddle operator.
    \item We propose EFTP as a training algorithm for GANs and show its convergence to a NE of the corresponding game, even in the regularized case.
\end{itemize}

We validate our analysis with some numerical simulations showing the effects of our proposed regularization in stabilizing the training process and the convergence properties of the proposed algorithm. 
Although our work takes GANs as a primary application, our results can be generalized to symmetric regularization, hence retrieving, for instance, the classic Tikhonov \cite{Tikhonov1995,fabiani2022stochastic}. As a consequence, the extension to $n$-players games is also possible, as long as some distinction between the two sides (generator and discriminator) is kept for the asymmetric case.
\subsection{Notation}
\label{subsec:notation}

Let $\RR$ be the set of all real number and $\RR^n$ represents $n$-dimensional Euclidean space.

\paragraph*{Probability}
Let $(X,\mathcal{X},p_{\mathcal D})$ and $(Z,\mathcal{Z},p_Z)$ be probability spaces.  
$X$ and $Z$ are the domains of data and latent vectors, respectively (e.g., $X=\mathbb{R}^{d_x}$, $Z=\mathbb{R}^{d_z}$).  $\mathcal X$ and $\mathcal Z$ are their $\sigma$-algebras, 
and $p_{\mathcal D}$ and $p_Z$ are probability measures on $(X,\mathcal X)$ and $(Z,\mathcal Z)$, respectively. We write $x\sim p_{\mathcal D}$ and $z\sim p_Z$ for samples drawn from these distributions.

\paragraph*{Matrices}


Given $A\in\mathbb{R}^{n\times n}$, we write \(\lambda_{\min}(A)\) and \(\lambda_{\max}(A)\) for its smallest and largest eigenvalues, respectively.
Let $\mathbb{M}^n\subset\RR^{n\times n}$ be the set of symmetric matrices of dimension $n$.
For $A\in\mathbb{M}^n$:
\(A\succeq 0\) or \(A\in\mathbb{M}^n_+\) means that A is positive semidefinite (PSD), i.e., \(x^\top A x \ge 0\) for all \(x\in\mathbb{R}^n\); 
\(A\succ 0\) or \(A\in\mathbb{M}^n_{++}\)  means that A is positive definite (PD), i.e., \(x^\top A x > 0\) for all \(x\neq 0\).
For vectors, \(\|\cdot\|_2\) denotes the Euclidean norm.
For matrices \(A\in\mathbb{R}^{m\times n}\), \(\|A\|\) denotes the spectral norm induced by \(\|\cdot\|_2\), i.e.
$
\|A\|=\sup_{x\neq 0}\frac{\|Ax\|_2}{\|x\|_2}=\sigma_{\max}(A)=\sqrt{\lambda_{\max}(A^\top A)}.
$
If \(A\in\mathbb{M}\), \(\|A\|=\max_i |\lambda_i(A)|\); if \(A\in\mathbb{M}^n_{++}\), \(\|A\|=\lambda_{\max}(A)\).

\paragraph*{Operator Theory}
For a nonempty closed convex set $S$, the Euclidean projector is
$\Proj_S(x):=\arg\min_{y\in S}\|y-x\|_2$. The normal cone is
$N_S(x):=\{v:\langle v, y-x\rangle\le 0,\ \forall y\in S\}$ when $x\in S$, and empty otherwise.
A function $f:\RR^n\to\RR$ is $\mu$-strongly convex (concave) if $f-\frac{\mu}{2}\|\cdot\|^2$
($-f-\frac{\mu}{2}\|\cdot\|^2$), $\mu>0$, is convex.
A mapping $F:\RR^n\to\RR^n$ is: $L$-Lipschitz continous on $S$ if $\|F(u)-F(v)\|\le L\|u-v\|$ for all $u,v\in S$ with $L>0$; \emph{monotone} if $\langle F(u)-F(v),\,u-v\rangle\ge 0$ for all $u,v\in S$; \emph{$\mu$-strongly monotone} if $\langle F(u)-F(v),\,u-v\rangle\ge \mu\|u-v\|^2$ with $\mu>0$.
Given $P\in\Mplus$, the residual is
$R_P(\omega)\;:=\;P^{-1}(\omega-\Proj_{S}(\omega-P\,F(\omega))).$

For a mapping $F:\RR^{d}\!\to\!\RR^{m}$, we indicate with $JF\in\RR^{m\times d}$ the Jacobian.
For a twice continuously differentiable function $f:\mathbb{R}^d \to \mathbb{R}$ we indicate with $\nabla^2_{uu} f \in \mathbb{M}$ the Hessian and with $\nabla^2_{xy} f$ the mixed partial derivatives.





\section{Problem Description}
\label{sec:problem}


We model the training of GANs as a two-player zero-sum game between
the \emph{generator} $G_\theta: Z\!\to\! X$ (player~G) and the \emph{discriminator} $D_\varphi: X\!\to\!\mathbb{R}$ (player~D),
with parameters $\theta\in\Theta\subset\mathbb{R}^{d_\theta}$ and $\varphi\in\Phi\subset\mathbb{R}^{d_\varphi}$.
The strategy sets $\Theta$ and $\Phi$ capture architectural/regularization constraints
We write $S:=\Theta\times\Phi$ and $\omega=(\theta,\varphi)$.
The two players min--max optimisation problem reads as
\begin{multline}
\min_{\theta\in\Theta}\,\max_{\varphi\in\Phi}\;
\mathcal{L}(\theta,\varphi)
:=\E_{x\sim p_{\mathcal D}}\!\big[\Psi(D_\varphi(x))\big]\\
+\E_{z\sim p_Z}\!\big[\Psi(-D_\varphi(G_\theta(z)))\big],
\label{eq:minmax}
\end{multline}
therefore this problem is a zero-sum game.
A popular choice for the objective is $\Psi(t)=-\log(1+e^{-t})$ but other choices can be considered, as long as the following assumptions (Assumption \ref{ass:curv} in particular) are satisfied.

Accordingly, a pair $(\theta^\star,\varphi^\star)\in S$ is a Nash equilibrium iff
\begin{equation}\label{Nash equili}
\mathcal{L}(\theta^\star,\varphi)\ \le\ \mathcal{L}(\theta^\star,\varphi^\star)\ \le\ \mathcal{L}(\theta,\varphi^\star),
\qquad \forall\,(\theta,\varphi)\in S.
\end{equation}

The min-max problem in \eqref{eq:minmax} can be rewritten as a saddle problem and solved by means of a VI. To this aim, let the saddle operator associated to problem \eqref{eq:minmax} be
\begin{equation}
F(\omega)=
\begin{bmatrix}
\nabla_\theta \mathcal{L}(\theta,\varphi)\\[.2em]
-\,\nabla_\varphi \mathcal{L}(\theta,\varphi)
\end{bmatrix}.
\label{eq:F}
\end{equation}
Then, the equilibria of the game in \eqref{eq:minmax} can be characterized, thanks to the operator in \eqref{eq:F}, as the solutions of the $\mathrm{VI}(S,F)$
\begin{equation} 
\quad \text{find} \ \omega^\star \in S \ \text{s.t.}\
\ip{F(\omega^\star)}{\omega-\omega^\star}\ge 0,
\quad \forall\,\omega\in S.
\label{eq:VI}
\end{equation}

We next state the standard conditions ensuring well-posedness of \eqref{eq:minmax} and \eqref{eq:VI}.

\begin{assumption}[Local Sets]\label{ass:set}
$\Theta$ and $\Phi$ are nonempty, closed, and convex.  
\end{assumption}

    Note that, under Assumption \ref{ass:set}, the Euclidean projector $\Proj_{S}$, used in the later sections, is firmly non-expansive \cite[sec 3.1]{RyuBoydPrimer}.

\begin{assumption}[Smoothness of the base operator]\label{ass:L0}
The saddle operator $F$ in \eqref{eq:F} is $L_0$-Lipschitz continous on $S$.
\end{assumption}

\begin{assumption}[Curvature of the saddle objective]\label{ass:curv}

The objective function $\mathcal{L}$ in \eqref{eq:minmax} is 
$\mu_\theta$-strongly convex in $\theta$ and 
$\mu_\varphi$-strongly concave in $\varphi$ on $S$, 
with moduli $\mu_\theta,\mu_\varphi\ge0$ 
\end{assumption}

\begin{remark}
    Note that in Assumption \ref{ass:curv} the parameters $\mu_\theta$ and $\mu_\varphi$ may be zero. This means that the objective function might be merely convex/concave (not strongly convex/strongly concave). Consequently, the symmetric part of the Jacobian JF is only positive semidefinite and the saddle operator $F$ is, at best, monotone on $S$.
\end{remark}

\subsection{Penalty Design}
\label{subsec:penalty}

Without regularization, the GAN saddle field is typically rotational and ill-conditioned, therefore projected first-order methods tend to oscillate or require impractically small steps sizes\cite{VIperspective,MeschederConverge,Mescheder2018Which,NagarajanKolterGAN}. To address these problems, we propose a regularization that injects curvature to suppress rotational drift and improve conditioning. The resulting field comes with explicit Lipschitz and (strong) monotonicity constants, which guide stepsize choices and enable convergence guarantees.  
Combining the base operator in \eqref{eq:F}  with a penalty produces
\begin{equation}
F_\gamma(\omega)=
\begin{bmatrix}
\nabla_\theta \mathcal{L}(\theta,\varphi)-\nabla_\theta B(\theta,\varphi)\\[.2em]
-\,\nabla_\varphi \mathcal{L}(\theta,\varphi)+\nabla_\varphi B(\theta,\varphi)
\end{bmatrix}.
\label{eq:Fgamma}
\end{equation}

We have a general, symmentric notation in \eqref{eq:Fgamma}, but later we propose an asymmetric one with $\nabla B_\theta(\theta,\varphi)=0$. Therefore, we regularize only the discriminator side of the game and analyze the induced operator $F_\gamma$.

\begin{remark}

Solving the regularized problem is not equivalent to solving the original one in general. In particular, solutions of $\mathrm{VI}(S,F_\gamma)$ need not coincide with those of $\mathrm{VI}(S,F)$.  
However, if $B$ vanishes at a saddle point of the unregularized operator, then $F_\gamma(\omega^\star)=F(\omega^\star)$.  
Moreover, as $\gamma\rightarrow0$, bounded solution sequences of $\mathrm{VI}(S,F_\gamma)$ have cluster points that solve $\mathrm{VI}(S,F)$.

To see this, let $\{\gamma_k\}_{k>0} \subseteq \mathbb{R}$ be a sequence of regularization parameters and let $\{\omega_k\}_{k>0}$ be the corresponding sequence of solutions of $\mathrm{VI}(S,F_{\gamma_k})$. Assume
$\{\omega_k\}$ is bounded with $\omega_k\to\bar\omega$ along a subsequence. For any
$y\in S$ we have $\langle F_{\gamma_k}(\omega_k),\,y-\omega_k\rangle\ge 0$ and by
uniform convergence $F_{\gamma_k}\to F$ on a compact set containing $\{\omega_k\}$ \cite[Thm.~4.3]{Shapiro2005SensitivityVI}. Therefore,
passing to the limit results in $\langle F(\bar\omega),\,y-\bar\omega\rangle\ge 0$ for all
$y\in S$, i.e., $\bar\omega \text{ is a solution of } \mathrm{VI}(S,F)$ \cite[Cor.~5.1.8]{facchinei2003finite}.
\end{remark}

We consider two forms of regularization. The terms are added in the cost and later we differentiate to obtain the gradients of the regularization terms in \eqref{eq:Fgamma}.

\paragraph{Tikhonov regularization (Tik)} Inspired by \cite{Tikhonov1995}, let
\begin{equation}
B_{\mathrm{Tik}}(\theta,\varphi)=\tfrac{\gamma}{2}\,\|\varphi\|^2.
\label{eq:Btik}
\end{equation}
Compared to the standard Tikhonov regularization, in our case the term acts only on the discriminator block. Accordingly,
$
\nabla_\varphi B_{\mathrm{Tik}}=\gamma\,\varphi,
\nabla^2_{\varphi\varphi}B_{\mathrm{Tik}}=\gamma I,
\nabla_\theta B_{\mathrm{Tik}}\equiv 0.
$
This penalty introduces curvature in the $\varphi$-subspace while leaving the generator block unaffected.

Moreover, if we also regularize the generator with the same quadratic term, i.e.,
\begin{equation}\label{eq:fulltik}
B_{\mathrm{Tik}}^{\mathrm{full}}(\theta,\varphi)
= \tfrac{\gamma}{2}(\|\theta\|^{2}+\|\varphi\|^{2}),
\end{equation}
we recover the \emph{classic (symmetric) Tikhonov} regularization and the penalty adds identical curvature to both the generator and discriminator subspaces.

\paragraph{Symmetric (zero-centered) Gradient Penalty (SGP)} As an alternative to the regularization introduced above, let

\begin{equation}\label{eq:Bsgp}
\begin{aligned}
B_{\mathrm{SGP}}(\theta,\varphi)
&= \tfrac{\gamma}{2}\,\mathbb{E}_{x\sim p_D}\!\big[\|\nabla_x D_\varphi(x)\|^2\big] \\
&\quad + \tfrac{\gamma}{2}\,\mathbb{E}_{z\sim p_Z}\!\big[\|\nabla_x D_\varphi(G_\theta(z))\|^2\big].
\end{aligned}
\end{equation}
This regularization penalizes the \emph{input gradient} of $D_\varphi$ toward zero over both real $(x)$ and generated $(z)$ supports. Despite its name\cite{Mescheder2018Which}, it is added exclusively to the discriminator payoff.  
By letting
$ g(x;\varphi):=\nabla_x D_\varphi(x)$ and $H_\varphi(x):=\nabla^2_{x,\varphi}D_\varphi(x),$
and by direct differentiation,
\begin{equation}\label{eq:extraBSGP}
\begin{aligned}
\nabla_\varphi B_{\mathrm{SGP}}
&= \gamma(
\mathbb{E}_{x\sim p_D}[ H_\varphi(x)^\top g(x;\varphi)] \\
&\quad +\, \mathbb{E}_{z\sim p_Z}[ H_\varphi(G_\theta(z))^\top g(G_\theta(z);\varphi)]).
\end{aligned}
\end{equation}
Using the chain rule through $x'=G_\theta(z)$ gives
$
\nabla_\theta B_{\mathrm{SGP}}
=\gamma\,\E\!\big[
J_\theta G_\theta(z)^\top
H_{xx}(x')^\top
g(x';\varphi)
\big],
$
where $H_{xx}(x):=\nabla^2_x D_\varphi(x)$ and $J_\theta G_\theta(z)$ is the generator Jacobian. Note that, although the regularization is added only on the discriminator cost, it affects also the generator blocks.



\begin{remark}
The gradients in \eqref{eq:Bsgp} are with respect to the input $x$, and not with respect to the parameter $\varphi$. This choice targets the function $D_\varphi(\cdot)$ in the data space: it penalizes sharpness of $D_\varphi$ on the real and generated
supports and is zero whenever $D_\varphi$ is locally flat there. Consequently, the penalty vanishes at the equilibrium and preserves the target saddle point \cite[Rem.~D.8]{Mescheder2018Which}. 
In particular, since $g(\cdot;\varphi^\star)\equiv 0$ at the equilibrium, \eqref{eq:Bsgp} vanishes at the saddle point, thereby preserving the equilibrium in \eqref{Nash equili}. This motivates the term "zero-centered", since both its value and its first derivative vanish at $(\theta^\star,\varphi^\star)$.
Compared with a parameter penalty (e.g., Tikhonov in \eqref{eq:Btik}), the input-gradient penalty acts in data space, it
is invariant to simple reparameterizations, and, since it injects curvature away from the saddle point, it does not bias the equilibrium, i.e., $\gamma$ need not be vanishing.

\end{remark}

\subsection{Regularization analysis}
To understand how the penalties modify the geometry of the saddle field, we examine the curvature they induce on the discriminator block. 
Since the local behavior of the saddle operator is governed by its Jacobian, to certify explicit Lipschitz continuity and strong–monotonicity constants for the regularized mapping, we inspect the second-order structure of the proposed penalties.

For the curvature analysis on the discriminator block, let the Gauss–Newton (GN) surrogate\cite[Sec.~10.3]{NocedalWright} of the Hessian be $\nabla^2_{\varphi\varphi}B_{\mathrm{SGP}}
\approx \gamma\,J_G(\omega)$, where 
\begin{equation}
J_G(\omega)
:=\E\!\big[
H_\varphi(x)^\top H_\varphi(x)
+H_\varphi(x')^\top H_\varphi(x')
\big]
\label{eq:JG}
\end{equation}
with $x$ being the real data inputs and $x'$ being the generated samples inputs.

\begin{assumption}[Gauss--Newton identifiability] 
There exist a neighborhood $\mathcal{R}\subseteq S$ and a constant $\lambda_0>0$ such that
$
J_G(\omega)\succeq \lambda_0 I,$ for all $\omega\in\mathcal{R}.
$
\label{ass:gn}
\end{assumption}
\begin{remark}
The Jacobian $JF_\gamma$ of $F_\gamma$ in \eqref{eq:Fgamma} includes off-diagonal mixed blocks stemming from $\nabla^2_{\theta\varphi}(\cdot)$ terms.  
These drive rotational dynamics and can 
affect the desired strong-monotonicity needed for our iterative scheme.  
Assumption~\ref{ass:gn} enforces discriminator-side curvature guaranteeing that the $\varphi$-block of the symmetric part of the Jacobian has a uniform positive bound (see Lemma \ref{lem:gramian}) that offsets the rotational coupling between generator and discriminator.
\end{remark}

\begin{assumption}[Continuity]\label{ass:C2}
The discriminator \(D_\varphi\) is twice continuously differentiable in both \(x\) and \(\varphi\) and on the relevant supports of \(x\sim p_D\) and \(x' = G_\theta(z)\), \(z\sim p_Z\).
\end{assumption}

Let \(\xi\) be a generic input to indicate either a real sample \(x\sim p_{\mathcal D}\) or a generated sample
\(x' = G_\theta(z)\) with \(z\sim p_Z\).



\begin{assumption}[Bounded mixed Hessian]\label{ass:Hbound}
There exists a constant \(C_H>0\) such that \(\|H_\varphi(\xi)\|\le C_H\) for all \(\xi\in\operatorname{supp}(p_{\mathcal D}) \cup G_\theta\big(\operatorname{supp}(p_Z)\big)\),
where \(H_\varphi(\xi):=\nabla^2_{x,\varphi}D_\varphi(\xi)\).
\end{assumption}
As a consequence
\(\mathbb E[\|H_\varphi(x)\|^2]\le C_H^2\) and
\(\mathbb E[\|H_\varphi(x')\|^2]\le C_H^2\).

\begin{lemma}[PSD and uniform bound for the Gramian]\label{lem:gramian}
Let Assumptions~\ref{ass:C2}--\ref{ass:Hbound} hold. 
Then \(J_G(\omega)\succeq 0\) and
\begin{equation}\label{eq_lemma}
\|J_G(\omega)\|\leq2C_H^2=:\kappa.
\end{equation}
\end{lemma}
\begin{proof}
Using the definition of $J_G(\omega)$ in \eqref{eq:JG}, to show that it is PSD, let $v\in\mathbb{R}^{d_\varphi}$; then
$
v^\top J_G(\omega)\,v
=\E\!\big[\|H_\varphi(x)v\|^2+\|H_\varphi(x')v\|^2\big]\ge 0,
$
hence $J_G(\omega)\succeq 0$.

The bound in \eqref{eq_lemma} instead follows by using linearity of expectation and the triangle inequality:
\begin{align*}
\|J_G(\omega)\|
&=\|\E[H_\varphi(x)^\top H_\varphi(x)]
      +\E[H_\varphi(x')^\top H_\varphi(x')]\| \\
&\le\;\|\E[H_\varphi(x)^\top H_\varphi(x)]\|
      +\|\E[H_\varphi(x')^\top H_\varphi(x')]\| \\
&\le\;\E\![\|H_\varphi(x)^\top H_\varphi(x)\|]
   +\E\![\|H_\varphi(x')^\top H_\varphi(x')\|].
\end{align*}
Since the spectral norm is convex, by Jensen inequality, 
$\|J_G(\omega)\|
\le\E[\|H_\varphi(x)\|^2]+\E[\|H_\varphi(x')\|^2].$
Finally by Assumption \ref{ass:Hbound}
$
\|J_G(\omega)\|\le C_H^2+C_H^2=2C_H^2=:\kappa.
$
which proves the claim.
\end{proof}

We can now show Lipschitz continuity and strong monotonicity of the reguralized operator in \eqref{eq:Fgamma}.

\begin{proposition}[Lipschitz constant of $F_\gamma$]\label{prop:Lipschitz}
Under Assumptions~\ref{ass:set}, \ref{ass:L0}, \ref{ass:C2} and \ref{ass:Hbound}, $F_\gamma$ in \eqref{eq:Fgamma} is $L$-Lipschitz continous on $S$ with
\begin{equation}
L\;=\;L_0+\gamma\,\kappa_{\rm tot},\qquad \kappa_{\rm tot}\in[\kappa,\;\kappa+\tilde\kappa_\theta].
\label{eq:L}
\end{equation}
where $\kappa$ is as in \eqref{eq_lemma},  and 
$
\tilde\kappa_\theta\ge 0.
$
\end{proposition}

\begin{proof}
By Assumption~\ref{ass:L0}, the operator $F$ is $L_0$-Lipschitz on $S$.
Therefore, it suffices to bound the Lipschitz constant of the regularization contribution
$\nabla B(\omega) = (\,-\nabla_\theta B(\omega),\,\nabla_\varphi B(\omega)\,)^T$.

\noindent\emph{(i) Discriminator-side contribution.}
For $B_{\mathrm{Tik}}$ in \eqref{eq:Btik}, we have $\nabla_\varphi B_{\mathrm{Tik}}(\omega)=\gamma\,\varphi$, whose Jacobian w.r.t.\ $\varphi$ is $\gamma I$; thus
\[
\|\nabla_\varphi B_{\mathrm{Tik}}(\omega)-\nabla_\varphi B_{\mathrm{Tik}}(\omega')\|
\;\le\; \gamma\,\|\omega-\omega'\|.
\]
For $B_{\mathrm{SGP}}$ in \eqref{eq:extraBSGP},
by the Gauss--Newton surrogate in \eqref{eq:JG} and Lemma~\ref{lem:gramian},
the $\varphi$-Jacobian of $\nabla_\varphi B_{\mathrm{SGP}}$ satisfies
$
\left\|\nabla^2_{\varphi\varphi} B_{\mathrm{SGP}}(\omega)\right\|
\;\approx\; \gamma\,\|J_G(\omega)\| \;\le\; \gamma\,\kappa,
$
with $\kappa$ as in \eqref{eq_lemma}. By the mean-value theorem for vector fields,
\[
\|\nabla_\varphi B_{\mathrm{SGP}}(\omega)-\nabla_\varphi B_{\mathrm{SGP}}(\omega')\|
\;\le\; \gamma\,\kappa\,\|\omega-\omega'\|.
\]
Thus, in both $B_{\mathrm{Tik}}$ and $B_{\mathrm{SGP}}$ cases, the discriminator-side contribution is $\gamma\,\kappa$-Lipschitz (with $\kappa=1$ for $B_{\mathrm{Tik}}$ and $\kappa$ from \eqref{eq_lemma} for $B_{\mathrm{SGP}}$).

\noindent\emph{(ii) Generator-side contribution.}
For $B_{\mathrm{Tik}}$ in \eqref{eq:Btik} we have $\nabla_\theta B_{\mathrm{Tik}}\equiv 0$, hence $\tilde\kappa_\theta=0$. For $B_{\mathrm{SGP}}$ in \eqref{eq:Bsgp} we have
$
\nabla_\theta B_{\mathrm{SGP}}(\omega)
=
\gamma\,\mathbb{E}\!\left[J_\theta G_\theta(z)^\top\,H_{xx}(x')^\top\,g(x';\varphi)\right]
$.
Under Assumptions~\ref{ass:C2}--\ref{ass:Hbound} and bounded supports/parameters, 
$J_\theta G_\theta$, $H_{xx}$ and $g$ are uniformly bounded and Lipschitz continuous (see Remark 6). Therefore,
$\nabla_\theta B_{\mathrm{SGP}}$ is Lipschitz continuous with some finite constant, i.e., $\tilde\kappa_\theta<\infty$.

Combining (i) and (ii), the regularization term is $\gamma(\kappa+\tilde\kappa_\theta)$-Lipschitz continuous. We have
$
\|F_\gamma(\omega)-F_\gamma(\omega')\|
\;\le\; \big(L_0+\gamma\,\kappa+\gamma\,\tilde\kappa_\theta\big)\,\|\omega-\omega'\|.
$
Since $\tilde\kappa_\theta\ge 0$, this yields \eqref{eq:L} with
$\kappa_{\rm tot}\in[\kappa,\;\kappa+\tilde\kappa_\theta]$.
\end{proof}

\begin{algorithm}[t]
\caption{Forward--Backward }
\label{alg:fb}
\begin{algorithmic}[1]
\Require initial $\omega_0\in \mathcal R$, stepsize/preconditioner $P\succ 0$
\For{$k=0,1,2,\dots$}
  \State $\omega_{k+1}\gets \Proj_{\mathcal{R}}\!\big(\omega_k-P\,F_\gamma(\omega_k)\big)$
\EndFor
\end{algorithmic}
\end{algorithm}

\begin{algorithm}[t]
\caption{ExtraGradient }
\label{alg:eg}
\begin{algorithmic}[1]
\Require initial $\omega_0\in \mathcal R$, stepsize/preconditioner $P\succ 0$
\For{$k=0,1,2,\dots$}
  \State $\tilde\omega_k\gets \Proj_{\mathcal{R}}\!\big(\omega_k-P\,F_\gamma(\omega_k)\big)$ \Comment{look-ahead}
  \State $\omega_{k+1}\gets \Proj_{\mathcal{R}}\!\big(\omega_k-P\,F_\gamma(\tilde\omega_k)\big)$
\EndFor
\end{algorithmic}
\end{algorithm}

\begin{algorithm}[t]
\caption{EFTP }
\label{alg:popov-correct}
\begin{algorithmic}[1]
\Require initial $\omega_0\in \mathcal R$, stepsize/preconditioner $P\succ 0$
\State $y_0\gets \omega_0$;\quad $\widehat F_0 \gets F_\gamma(y_0)$ \Comment{warm-start: one oracle call}
\For{$k=0,1,2,\dots$}
  \State $\omega_{k+1}\gets \Proj_{\mathcal{R}}\!\big(\omega_k - P\,\widehat F_k\big)$
  \State $y_{k+1}\gets \Proj_{\mathcal{R}}\!\big(\omega_{k+1} - P\,\widehat F_k\big)$
  \State $\widehat F_{k+1} \gets F_\gamma(y_{k+1})$ \Comment{\textbf{one} oracle call per loop}
\EndFor
\end{algorithmic}
\end{algorithm}

\begin{remark}
    The quantity $\tilde\kappa_\theta$ in \eqref{eq:L} is nonnegative by definition and it captures the $\theta$-dependence of $B$. In the case of the Tikhonov regularization in \eqref{eq:Btik}, $\tilde\kappa_\theta=0$. For the SGP regularization in \eqref{eq:Bsgp}, $\tilde\kappa_\theta>0$ whenever the generator Jacobian $J_\theta G_\theta$ and the mixed Hessian $H_\varphi$ are uniformly bounded on the supports of $(x,z)$. This happens when $x$ and $z$ range over bounded sets (e.g., $x\in[0,1]^{d_x}$, $z\in[-c,c]^{d_z}$), $G_\theta(Z)$ is bounded, the parameter sets $\Theta,\Phi$ are bounded, and $G_\theta$ and $D_\varphi$ are twice continuously differentiable on their domains.
\end{remark}


\begin{proposition}[Strong monotonicity of $F_\gamma$]\label{prop:strongmono}
Under Assumptions~\ref{ass:curv} and \ref{ass:gn}, the Jacobian $JF_\gamma$ of the regularized map $F_\gamma$ in \eqref{eq:Fgamma} satisfies
\begin{equation}
\frac{JF_\gamma(\omega)+JF_\gamma(\omega)^\top}{2}\ \succeq\ \mathrm{diag}\big(\mu_\theta I,\;(\mu_\varphi+\gamma\lambda_0)I\big)
\label{eq:sym}
\end{equation}
on $\mathcal R\subseteq S$. Therefore, $F_\gamma$ is $\mu$-strongly monotone on $\mathcal R$ with
\begin{equation}
\mu\;\ge\;\min\{\mu_\theta,\ \mu_\varphi+\gamma\lambda_0\}.
\label{eq:mu}
\end{equation}

\end{proposition}

\begin{proof}\label{app:proof-strongmono}
Let $F_\gamma$ be as in \eqref{eq:Fgamma}. Then, its Jacobian is
\begin{equation}
JF_\gamma(\omega)
=
\begin{bmatrix}
A & B_{\rm mix}\\[.2em]
-\,B_{\rm mix}^\top & C
\end{bmatrix}\omega,\\[.4em]
\label{eq:JFgamma}
\end{equation}
where $A=\nabla^2_{\theta\theta}\La-\nabla^2_{\theta\theta}B$, $B_{\rm mix}=\nabla^2_{\theta\varphi}\La-\nabla^2_{\theta\varphi}B$ and $C=-\,\nabla^2_{\varphi\varphi}\La+\nabla^2_{\varphi\varphi}B$.
By Schwarz symmetry, the \emph{symmetric} part is block diagonal:
$
\frac{JF_\gamma(\omega)+JF_\gamma(\omega)^\top}{2}
=\operatorname{diag}(
\nabla^2_{\theta\theta}\mathcal{L}-\nabla^2_{\theta\theta}B,\;
-\,\nabla^2_{\varphi\varphi}\mathcal{L}+\nabla^2_{\varphi\varphi}B
)
$.
By Assumption~\ref{ass:curv}, $\nabla^2_{\theta\theta}\La\succeq \mu_\theta I$ and $-\nabla^2_{\varphi\varphi}\La\succeq \mu_\varphi I$. We consider the two regularization separately.
\begin{enumerate}[label=\textit{\alph*)}]
\item For Tikhonov regularization, $\nabla^2_{\varphi\varphi}B_{\rm Tik}=\gamma I$ and $\nabla^2_{\theta\theta}B_{\rm Tik}\equiv 0$, hence on any region in $\mathcal{R}$, $\frac{JF_\gamma(\omega)+JF_\gamma(\omega)^\top}{2}\ \succeq\ \mathrm{diag}\big(\mu_\theta I,\;(\mu_\varphi+\gamma)I\big).
$
\item 
For $B_{\rm SGP}$, using the Gauss–Newton surrogate in \eqref{eq:JG} and Assumption~\ref{ass:gn}, $\nabla^2_{\varphi\varphi}B_{\rm SGP}\succeq \gamma\,J_G(\omega)\succeq \gamma\lambda_0 I$ on $\mathcal R$, hence 
$
\frac{JF_\gamma+JF_\gamma^\top}{2}\ \succeq\ \mathrm{diag}\big(\mu_\theta I,\;(\mu_\varphi+\gamma\lambda_0)I\big),
$
on $\mathcal R\subset S$ which is exactly \eqref{eq:sym}. 
\end{enumerate}
Therefore $F_\gamma$ is $\mu$-strongly monotone on $\mathcal R$ with
$\mu\ge \min\{\mu_\theta,\mu_\varphi+\gamma\lambda_0\}$ 
\cite[Proposition 2.3.2.(c)]{facchinei2003finite}. \qedhere
\end{proof}

\begin{remark}\label{rem:ensure-mu}
If $\mu_\theta=0$ in Assumption \ref{ass:curv}, Proposition 2 implies only $\mu\ge 0$ i.e., the operator $F_\gamma$ might be monotone. Adding generator-side curvature, e.g., a Tikhonov term $\frac{\gamma}{2}\|\theta\|^2$ as in \eqref{eq:fulltik}, the right-hand side of Equation \eqref{eq:sym} becomes $\mathrm{diag}\big((\mu_\theta+\tau)I,\ (\mu_\varphi+\gamma\lambda_0)I\big)$ and gives
$
\mu \;\ge\; \min\{\mu_\theta+\tau,\ \mu_\varphi+\gamma\lambda_0\}\;>\;0, 
$ retrieving strong monotonicity.
\end{remark}

\section{Algorithms and convergence}\label{sec:convergence}

We consider three projected first-order methods to find a Nash equilibrium of our saddle problem:
\emph{Forward--Backward (FB)}, \emph{ExtraGradient (EG)}, and \emph{Extrapolation from the Past (EFTP)}. The details of these iterative schemes are detailed in Algorithms 1, 2 and 3, respectively. Note that we use the regularized operator in \eqref{eq:Fgamma} for these iterative schemes.

FB is the classical projected gradient step, taking a single projection along \(-F_\gamma(\omega_k)\)\cite{FranciGrammatico2022}. 
EG stabilizes the dynamics by evaluating at a \emph{look-ahead} point \(\tilde{\omega}_k\) and updating with \(F_\gamma(\tilde{\omega}_k)\), which helps cancel rotational drift on monotone saddle fields \cite{Korpelevich1976}. 
EFTP achieves similar stabilization while reusing a stored evaluation: after a warm start with one oracle call, it needs only a single gradient call per iteration, reducing computational cost \cite{Popov1980AMO}.

\subsection{Convergence analysis}\label{subsec:stepsizes}

Let $L$ be the Lipschitz constant of $F_\gamma$ on $S$ and $\mu$ its (strong)-monotonicity
constant on the region $\mathcal R\subseteq S$ as in Lemma \ref{lem:gramian} and Proposition \ref{prop:Lipschitz}. We recall that, according to Remark \ref{rem:ensure-mu}, $\mu$ might be 0.

\begin{assumption}[Stepsizes]\label{ass:steps}
Let $\mathcal R\subseteq S$ and $P\in\mathbb{M}^{d}_{++}$, the step size sequences are such that
for FB, $\ \|P\|^2 <\frac{2\mu\lambda_{\min}(P)}{L^2}$; for EG, $\ \|P\| \le \frac{1}{L}$; for EFTP, $\ \|P\|\le\frac{1}{4L}\ $.
\end{assumption}

\begin{remark}[Scalar case]
If $P=\eta I_d$, then the bounds reduce to $0<\eta<\tfrac{2\mu}{L^2}$ (FB), $0<\eta\le\tfrac{1}{L}$ (EG), and $0<\eta\le\tfrac{1}{2L}$ (EFTP).
Using the explicit bound $L=L_0+\gamma\,\kappa_{\rm tot}$ from Proposition~\ref{prop:Lipschitz} produces a computable step-sizes.
\end{remark}

We are now ready to state the convergence result for our iterative schemes.

\begin{theorem}[FB vs.\ EG]\label{thm:fb-eg}
Let Assumptions~\ref{ass:set}, \ref{ass:L0} and \ref{ass:steps} hold. 
\begin{enumerate}
  \item  If $F_\gamma$ is strongly monotone ($\mu>0$) on $\mathcal R$, then Algorithm \ref{alg:fb} converges to a solution $\omega^\star$ of \eqref{eq:VI}. The convergence is $Q$-linear.
  \item If $F_\gamma$ is monotone ($\mu=0$) on $S$, then Algorithm \ref{alg:eg} converges to a solution $\omega^\star$ of \eqref{eq:VI}. If, in addition, $F_\gamma$ is \emph{strongly monotone} on $\mathcal R$, the convergence is $Q$-linear.
\end{enumerate}
\end{theorem}

\begin{proof}
1)
Consider the fixed-point map
$T(\omega)=\Pi_S\!\big(\omega-P^{-1}F_\gamma(\omega)\big)$.
By \cite[Thm.~12.1.2]{facchinei2003finite}, and since $F_\gamma$ is $\mu$-strongly monotone and $L$-Lipschitz continuous on $\mathcal R$, $T$ is a contraction in the $P^{-1}$-induced norm. 
Therefore, $\omega_{k+1}=T(\omega_k)$ converges to the unique solution of $\mathrm{VI}(S,F_\gamma)$,
and the Banach contraction theorem provides $Q$-linear convergence.

2)
If $F_\gamma$ is monotone on $S$ and $L$-Lipschitz continuous, Algorithm \ref{alg:eg} generates a Fejér-monotone sequence with respect
to the solution set $\operatorname{SOL}(S,F_\gamma)$ and hence converges to a solution of
$\mathrm{VI}(S,F_\gamma)$\cite[Thm 12.1.1]{facchinei2003finite}, \cite[§5.8]{RyuBoydPrimer}. If, in addition, $F_\gamma$ is (locally) strongly
monotone on $\mathcal R$, then EG
converges $Q$-linearly for sufficiently small $\eta$
\cite[Thm.~12.6.4]{facchinei2003finite}.
\end{proof}

\begin{remark}[Region of validity $\mathcal R$]
Throughout Sections ~\ref{sec:problem}--\ref{sec:convergence}, 
$\mathcal R\subseteq S$ denotes a convex neighborhood of a solution 
$\omega^\star$ on which our regularity assumptions hold and where the
constants $L$ (Lipschitz continuity) and $\mu$ (strong monotonicity) are valid. 
All convergence statements for FB, EG, and EFTP are made on $\mathcal R$.
If $\mathcal R=S$, the results are global; otherwise they are local and
require an initial point $\omega_0\in\mathcal R$.
\end{remark}

We state the convergence of Algorithm 3 as a separate result since, as far as we know, it is a new scheme for games.

\begin{theorem}[EFTP]\label{thm:linear}
Let Assumptions \ref{ass:set}-\ref{ass:steps} hold. 
\begin{enumerate}
  \item  If $F_\gamma$ is monotone ($\mu=0$) on $\mathcal R$, then Algorithm \ref{alg:popov-correct} converges to a solution $\omega^\star$ of \eqref{eq:VI}. 
  \item If $F_\gamma$ is strongly monotone ($\mu>0$) on $R$, then Algorithm \ref{alg:popov-correct} is a contraction on $\mathcal R$ and
$
\|\omega_{k+1}-\omega^\star\|\leq\,\|\omega_k-\omega^\star\|$ with $q\in(0,1)$.
Consequently, Algorithm \ref{alg:popov-correct} converges $Q$-linearly to a solution $\omega^\star$ of \eqref{eq:VI}. Moreover, the number of iterations to reach $\|\omega_{k+1}-\omega^\star\|\le\varepsilon$ scales as $\mathcal O(\log(1/\varepsilon))$.
\end{enumerate}
\end{theorem}
\begin{proof}
    Let $\omega^\star\in \mathrm{SOL}(S,F_\gamma)$ be the (locally) unique solution in $\mathcal R$. Then, for 1), convergence follows mutatis mutandis from \cite[Theorem 1]{Popov1980AMO}. In particular, our conservative step size $\ \|P\|\le\frac{1}{4L}\ $ ensures the bound $\ \|P\|\, \,\le\tfrac{1}{3L}\ $ used in \cite{Popov1980AMO}.
    If, instead, we instead have strong monotonicity, analogously to \cite[Theorem'~1]{VIperspective} which shows 
    $\|\omega_{k+1}-\omega^\star\|\le\left(1-\frac{\mu}{4L}\right)\,\|\omega_k-\omega^\star\|$
    leading to $q\in(0,1)$ by Assumption \ref{ass:steps}. Accordingly, by \cite[Thm.~2.1.21(c)]{facchinei2003finite}, the number of iterations to reach $\|\omega_{k+1}-\omega^\star\|\le\varepsilon$ 
scales as $\mathcal{O}(\log(1/\varepsilon))$.
\end{proof}

\section{Numerical simulations}
\label{sec:bilinear}
\begin{figure}[t]
  \centering
  \includegraphics[width=.9\linewidth]{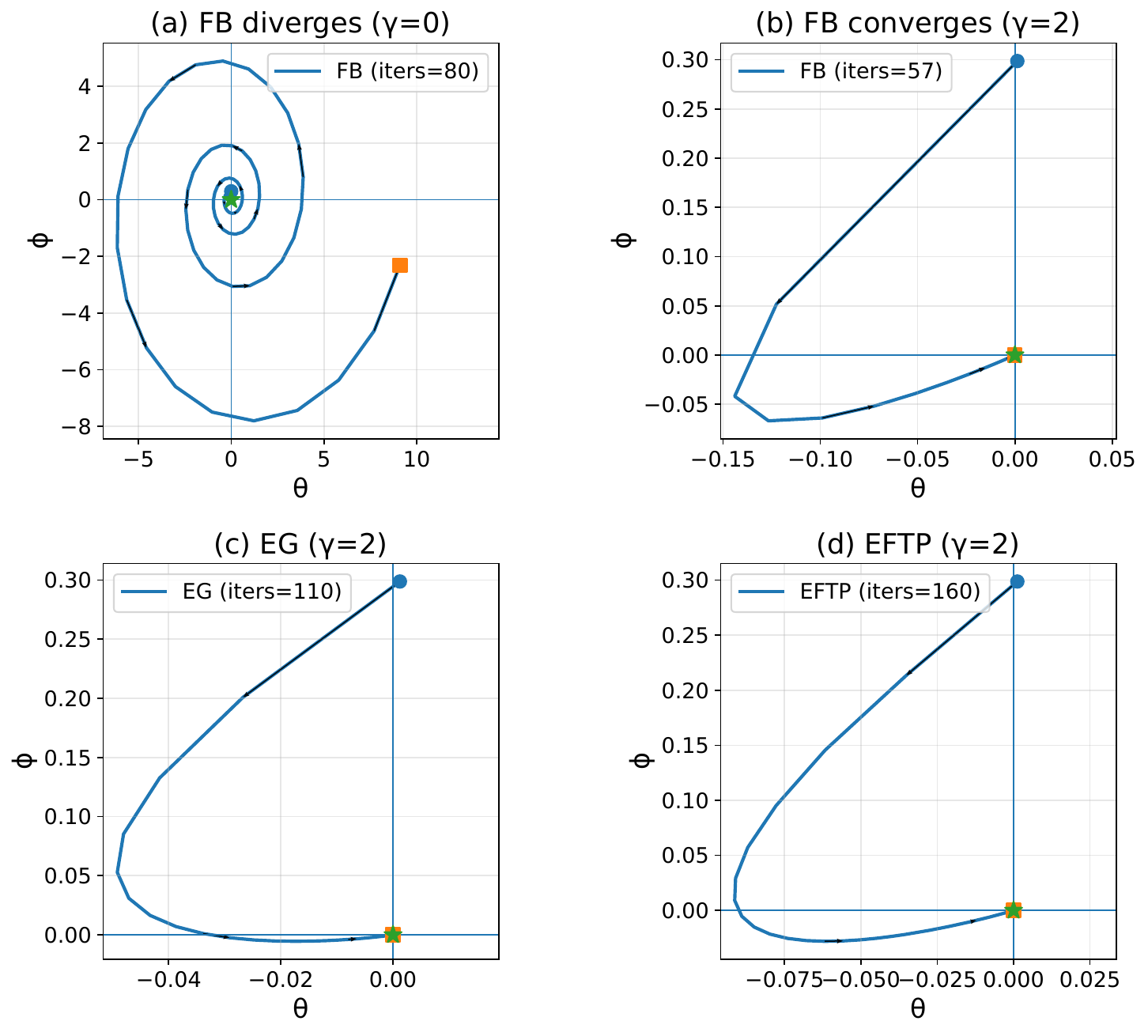}
  \caption{Trajectories with discriminator-only curvature: 
  (a) FB diverges for $\gamma=0$; 
  (b) FB converges for $\gamma=2$; 
  (c) EG ($\gamma=2$); 
  (d) EFTP ($\gamma=2$). 
  Blue dots indicate the starting point and squares the end points; the star is the solution. The legends report iteration till convergence.}
  \label{fig:bilinear-panel}
\end{figure}
To illustrate our asymmetric design mechanism, we study a bilinear toy example. 
We assume the discriminator is linear in the input, 
and specialize the GAN loss to
$\mathcal{L}(\theta,\varphi)= \mathbb{E}_{x\sim p_{\mathcal D}}[\varphi^\top x]
   - \mathbb{E}_{z\sim p_Z}[\varphi^\top G_\theta(z)] = \varphi^\top\big(\mathbb{E}[x]-\mathbb{E}[G_\theta(z)]\big).$
In particular, we set \(\mathbb{E}[x]=0\) and \(\mathbb{E}[G_\theta(z)]=-a\theta\) with \(a>0\). 
Since the discriminator is linear in its input, 
$g(x;\varphi)=\nabla_x D_\varphi(x)=\varphi$ is constant in $x$.  
In this case, the zero-centered input-gradient penalty simplifies to
\[
B_{\mathrm{SGP}}(\theta,\varphi)
=\tfrac{\gamma}{2}\,\E\!\big[\|g(x;\varphi)\|^2+\|g(G_\theta(z);\varphi)\|^2\big]
=\tfrac{\gamma}{2}\|\varphi\|^2,
\]
which exactly matches the Tikhonov regularization in \eqref{eq:Btik}. 
With \(\omega=(\theta,\varphi)^\top\in\RR^2\), the associated saddle operator is
\begin{equation}
\label{eq:bilinear-F}
F_\gamma(\omega)
=
\begin{bmatrix}
a\,\varphi \\[.15em]
-\,a\,\theta + \gamma\,\varphi
\end{bmatrix},
\qquad
J{F_\gamma}
=
\begin{bmatrix}
0 & a\\[.15em]
-\,a & \gamma
\end{bmatrix}.
\end{equation}

Note that $F_\gamma$ is \emph{monotone} for $\gamma\ge 0$, but not \emph{strongly} monotone\cite[Sec.~2.3]{facchinei2003finite}. We implemented \eqref{eq:bilinear-F} with $a>0$ and varied $\gamma\ge 0$. For stepsize we used Assumption \ref{ass:steps} with $L$ as in Proposition~\ref{prop:Lipschitz}.
Under this premises, 
FB exhibits oscillatory behavior and can stall unless $\gamma$ is large enough and the stepsize is tuned conservatively, i.e., according to Assumption \ref{ass:steps}. This can be observed in Figure~\ref{fig:bilinear-panel}(a), with $\gamma=0$ and Figure~\ref{fig:bilinear-panel}(b) with $\gamma=2$.
For EG and EFTP instead, 
the second gradient step cancels the rotational drift and results in convergence (Figure~\ref{fig:bilinear-panel}(c)-(d)).
Moreover, EFTP with one oracle call per half step matches the stabilizing effect of EG on monotone fields while using stored gradients hence while being computationally lighter. 

\begin{figure}[t]
  \centering
  \includegraphics[width=0.9\columnwidth]{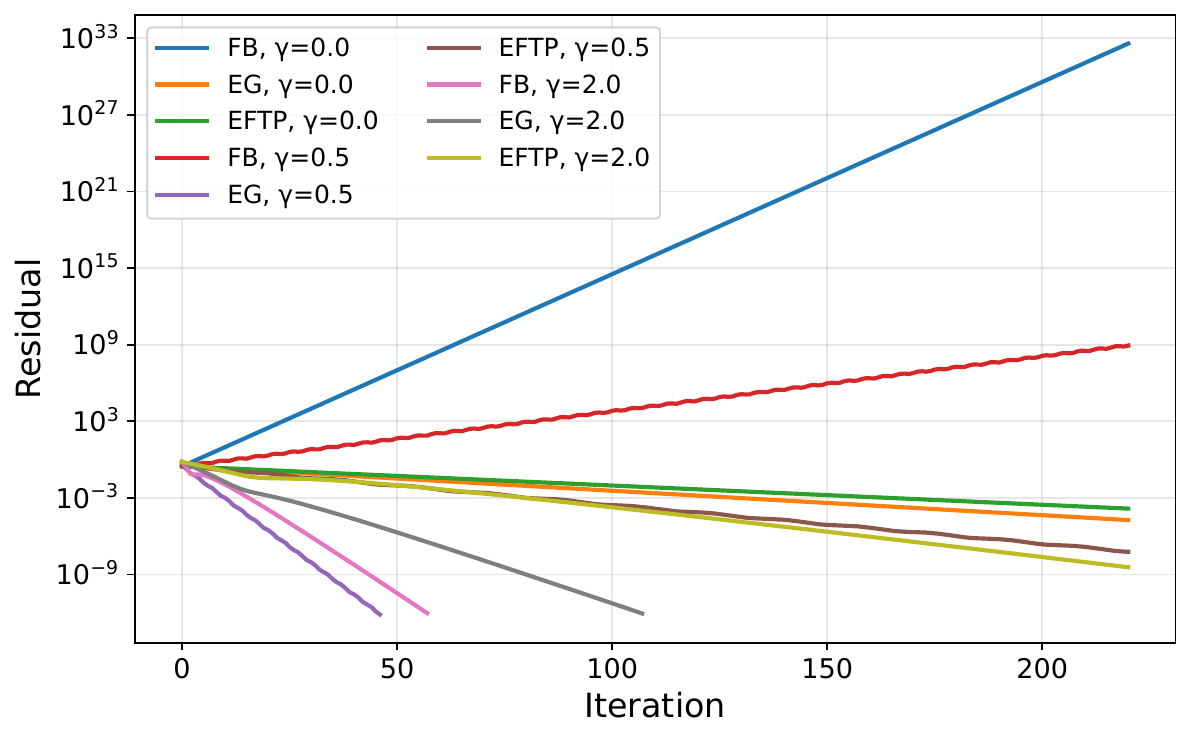}
  \caption{Residual vs. iterations on the bilinear toy for 
  $\gamma=\{0,0.5,2\}$. }
  \label{fig:residuals-iter}
\end{figure}
\begin{figure}[t]
  \centering
  \includegraphics[width=0.9\columnwidth]{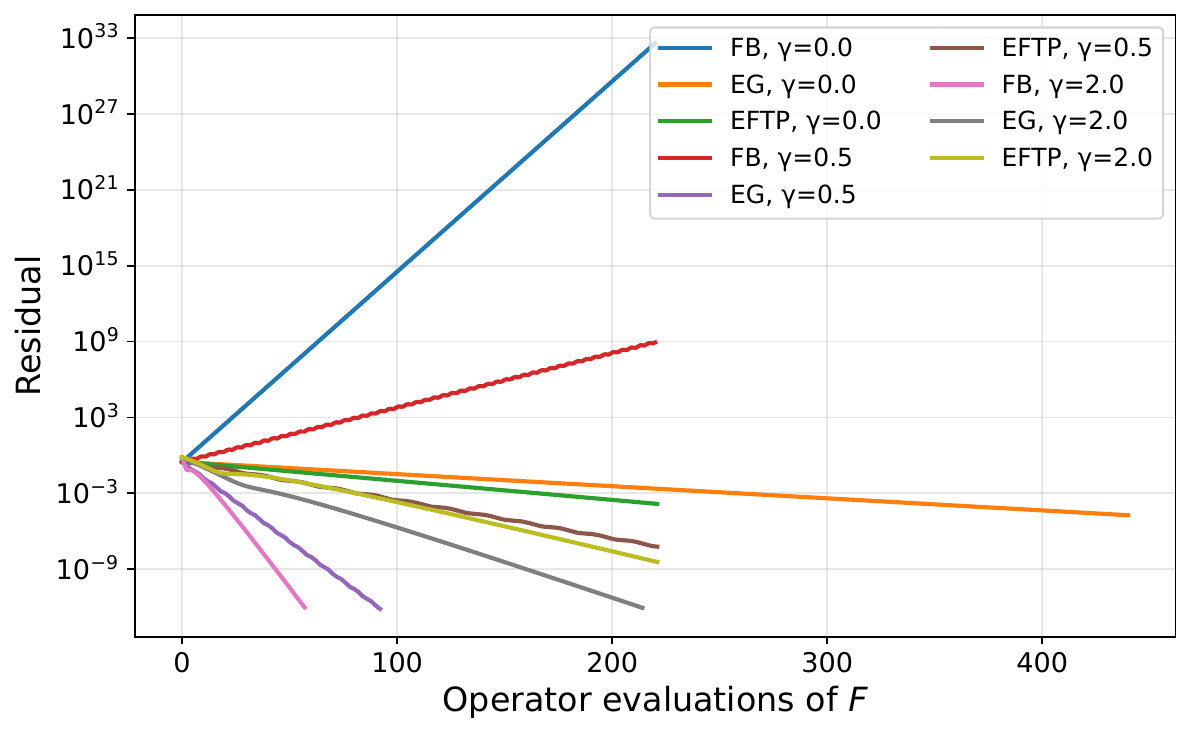}
  \caption{Residual as a function of gradient evaluations.}
  \label{fig:residuals-evals}
\end{figure}



The residual plots in Figures~\ref{fig:residuals-iter}–\ref{fig:residuals-evals} show how $\gamma$
stabilizes the adversarial dynamics: FB diverges for $\gamma=0$, slows for moderate curvature,
and contracts linearly for $\gamma=2$. For all $\gamma$, the field remains monotone but not strongly
monotone, so only look-ahead methods (EG, EFTP) achieve convergence. In particular, EG and EFTP exhibit stable, near-linear decrease that improves with larger $\gamma$, whereas FB is unstable for small curvature and stabilizes as $\gamma$ increases. Concerning the computational burden (Figure \ref{fig:residuals-evals}),
EG uses two gradient steps per iteration and EFTP only one per half step: both eliminate the rotational drift but EFTP achieves similar performances with fewer gradient calls.

\section{Conclusion}\label{sec:conclusion}

We studied GAN training through a VI formulation and introduced a block-asymmetric, zero-centered input-gradient penalty applied to the discriminator only. This design preserves the target saddle point while certifiably improving conditioning. Under smoothness and a local Gauss--Newton (GN) identifiability condition, we derived explicit Lipschitz continuity and strong-monotonicity constants for the regularized operator and established linear convergence of EFTP scheme. Extending the analysis to stochastic oracles with variance reduction or adaptive block preconditioners tasks would further test the asymmetric design.

\bibliographystyle{IEEEtran}
\bibliography{refs}

\end{document}